\documentclass[twocolumn,prb,aps,superscriptaddress]{revtex4}
\usepackage{amsmath,amssymb}
\usepackage[colorlinks, citecolor=blue]{hyperref}
\usepackage{amsfonts}
\usepackage{mathrsfs}
\usepackage{graphicx}
\usepackage{amssymb}
\usepackage{color}
\usepackage{dcolumn}
\usepackage{bm}
\usepackage[header,title,page,titletoc]{appendix}

\begin{document}

\title{Duality of deconfined quantum critical point in two
dimensional Dirac semimetals}
\author{Jiang Zhou}
\thanks{} \email{jzhou5@gzu.edu.cn}
\affiliation{Department of Physics, Guizhou University, Guiyang 550025, PR China}
\author{Ya-jie Wu}
\affiliation{School of Science, Xi'an Technological University, Xi'an 710032, PR China}
\author{Su-Peng Kou}
\thanks{} \email{spkou@bnu.edu.cn}
\affiliation{Department of Physics, Beijing Normal University, Beijing 100875, PR China}

\begin{abstract}
In this paper we discuss the N$\acute{e}$el and Kekul$\acute{e}$ valence bond
solids quantum criticality in graphene Dirac semimetal. Considering the
quartic four-fermion interaction $g(\bar{\psi}_i\Gamma_{ij}\psi_j)^2$ that contains
spin,valley, and sublattice degrees of freedom in the continuum field
theory, we find the microscopic symmetry is spontaneously broken when the
coupling $g$ is greater than a critical value $g_c$. The symmetry breaking
gaps out the fermion and leads to semimetal-insulator transition. All possible
quartic fermion-bilinear interactions give rise to the uniform critical
coupling, which exhibits the multicritical point for various orders and the
Landau-forbidden quantum critical point. We also investigate the typical
critical point between N$\acute{e}$el and Kekul$\acute{e}$ valence bond solid transition
when the symmetry is broken. The quantum criticality is captured by the
Wess-Zumino-Witten term and there exist a mutual-duality for N$\acute{e}$el-Kekul$\acute{e}$ VBS
order. We show the emergent spinon in the N$\acute{e}$el-Kekul$\acute{e}$ VBS transition , from which we conclude the phase
transition is a deconfined quantum critical point. Additionally, the connection between the index theorem and zero energy mode
bounded by the topological defect in the Kekul$\acute{e}$ VBS phase is studied to reveal the N$\acute{e}$el-Kekul$\acute{e}$ VBS duality.
\end{abstract}

\maketitle

\section{\textbf{Introduction}}

The study of quantum phases of matter and the phase transition in strongly
correlated systems is one of the central issues of modern condensed matter physics
\cite{sdev,qi0,hasan,fu,read}. In past seven years, the semimetal phases have
attracted extensive attentions. In topological Weyl/Dirac semimetals, the
conduction and valence bands cross at zero-dimensional nodal points\cite
{weng,weng1,weng2,wan,vish,xsy,lbq,hou}, but for topological nodal line
semimetals, it crosses at one-dimensional nodal line(see Refs.~\onlinecite{cw, wz} and
references therein). Recently, based on the 2D and 3D semimetal, the
fermion-boson mixed system was proposed to describe the Landau-forbidden
quantum phase transition, which gives rise to Fermion-induced
quantum critical points(FIQCP)\cite{yao,yao1,yao2,yyz} and fluctuation-induced
continuous phase transition with hidden degrees of freedom \cite{herbut1,herbut2,herbut3}. The FIQCP is type-II
Landau-forbidden transitions and the type-I is deconfined quantum critical
point in quantum magnet\cite{tsen}. In addition, the fermion-boson mixed
system have been shown to provide us well understanding of space-time
supersymmetry at quantum critical point(QCP)\cite{gro,lee,yao3}.

Graphene is a Dirac semimetal phase, and the low energy quasiparticle
excitations are described by (2+1) dimensions field theory of free Dirac
fermions. Due to the special relativistic dispersion, the graphenelike
structure provides a platform for realization of many quantum electrodynamics
predictions(e.g. Klein tunneling \cite{geim}, Zitterbewegung effect \cite{kat}
). When suffering from perturbations or electron correlation, the semimetal
opens a fermion gap and leads to rich phases of matter. For example, the
integer quantum Hall state is obtained by introducing the next-nearest
neighbor hopping with complex phases but without net magnetic flux threads each hexagonal plaquette
\cite{haldane}, $\mathbb{Z}_2$ quantum spin
Hall effect with time reversal symmetry could be obtained in a similar way \cite{kane}. The charge
density wave, Kekul$\acute{e}$ dimerization(Kekul$\acute{e}$ valence-bond-solid), and spin
density wave are known orders that gap out the fermion spectrum \cite
{kou1,cham}. Also, the fermi-boson coupling gaps out the fermion and leads to the surprising fractional charge \cite{cham,mudry,franz,jack}. It's
shown that the semimetal exhibits exotic topological phases\cite{xu1,xu2} and
the non-Landau-Ginzburg-Wilson transition\cite{yyz,herbut1,ass} with electron correlation and emergent fluctuation. The fluctuation with emergent degree of freedom  gives rise to the rich quantum criticality behavior and the various QCP is shown to belong to the Gross-Neveu-Yukawa(GNY) universality class\cite{herbut3,herbut4,herbut5,herbut6,gn}. Meanwhile, recent large-scale quantum Monte Carlo simulations have
confirmed that the quantum criticality in interacting Dirac semimetals is
consistent with the Gross-Neveu universality class\cite{ass1,ass2,yo}.

In this paper, by using the nonlinear sigma model field theory \cite
{tan,hx,tsen2,egm,yao4}, we understand the N$\acute{e}$el and Kekul$\acute{e}$
valence-bond-solid(VBS) quantum criticality in Dirac semimetals. Our motivation
is stated as follows. Recently, Ref.\onlinecite{yao} have studied the
semimetal-Kekul$\acute{e}$ VBS transition, and shown the evidence of a deconfined
quantum critical point between the Kekul$\acute{e}$-VBS and antiferromagnetic
phases of $SU(N)$ fermions model on the honeycomb lattice($N=2$ is the
flavors of four-component Dirac fermions). All the N$\acute{e}$el and Kekul$\acute{e}$ VBS are
gapped quantum phases, the induced mass-gap corresponds to the modulation of the fermion bilinear.
The coupling between the N$\acute{e}$el-Kekul$\acute{e}$ VBS fermion bilinear and the fermion fields leads to the purely bosonic fields theory for competing N$\acute{e}$el-Kekul$\acute{e}$ VBS orders. From the corresponding dynamical of bosonic fields, the quantum criticality will be easily identified. The
phase transition is shown to associate with a emergent spinon and gauge field at the critical point, so the
N$\acute{e}$el-Kekul$\acute{e}$ VBS transition indeed a deconfined QCP.

The mass-gap equivalent to the dynamical symmetry breaking-induced massive fermion and the
$(2+1)$ GNY theory is crucial for the mass-gap generation. In correlated many-body systems, we often encounter the Yukawa terms that driven by short range
fermion-fermion interacting.
For mass generation and competing phase in interacting systems,
the instability of interacting play a key role. Apart from the short range
Coulomb interaction \cite{gama}, the general four-body interaction in graphene semimetals includes
the quartic fermion bilinear that mixed the spin, valley, and sublattice. When
the symmetry is broken, the general fermion bilinear generates various mass
matrices that anticommute with each other. The physical explanation of these mass matrices have been
established in Ref.\onlinecite{ryu}. We will show there exists multicritical point for possible orders, and the multicritical point only
depend on the dimensions of the mass matrix. The emergence of two symmetry-unrelated orders at a critical point
display the Landau-beyond quantum criticality, and the effective action for
a set of anticommutating orders may be derived upon integrating over the fermion
fields \cite{weigman}. Under such formalism, the quantum criticality and the quantum phase transition between different phases can be
readily understood.

As shown in the text, the N$\acute{e}$el and Kekul$\acute{e}$ VBS orders at the critical point are dual to each other, and the topological excitation or topological defect also share the duality in two phases.
The topological defect bounds midgap fermion spectrum(i.e. zero energy mode) in the core is well known for fermi-boson mixed field theory.
Jackiw and Rossi have studied
the fermion-vortex system with fermion-scalar field interaction, and point
that the Dirac equation possesses $|n|$ zero-eigenvalue modes in the $n$
-vorticity background field \cite{jack2}. This result has been proven
associate with a index theorem\cite{weinberg}. The fermion zero mode has caused great interests in various condensed matter
systems \cite{gura,cheng,jack3}. Here, we study the zero energy mode bounded by the topological texture in Kekul$\acute{e}$ VBS phase as a complementary study.
And more importantly, we study the index of the topological excitation beyond any symmetries in the sense of N$\acute{e}$el-Kekul$\acute{e}$ VBS duality.

The paper is organized as follows. The dynamical gap equation is established
in sec. II. From the gap equation, we solve uniform critical coupling
constant for possible broken orders. In Sec. III, the typical N$\acute{e}$el-Kekul$\acute{e}$
VBS transition is studied. We derive the Wess-Zumino-Witten action for
five-component N$\acute{e}$el-Kekul$\acute{e}$ VBS order vector and discuss the N$\acute{e}$el-Kekul$\acute{e}$ VBS
mutual duality. In Sec. IV, we study the index theorem of zero energy mode
in Kekul$\acute{e}$ VBS state to reveal the N$\acute{e}$el-Kekul$\acute{e}$ VBS mutual duality. In the
end, we give a summary in Sec. V.

\section{gap generation and symmetry }

The hexagonal lattice of graphene semimetal contains two sets of triangular
sublattices, which denoted by $a$ and $b$. Near two inequivalent Dirac points ($\mathbf{k},\mathbf{k'}$) of
momentum space, The kinetic part of spinless graphene is described by the
Hamiltonian $H=\sum_{k}\psi _{k}^{\dag }h(k)\psi_{k}$, where
the fermion annihilation operators $\psi_{k}$ around the Dirac nodes is given by
\begin{equation}
\psi _{k}=\left( c_{\mathbf{k}a},c_{\mathbf{k'}b},c_{\mathbf{k'}b},c_{\mathbf{k}a}\right) ^{t}.
\end{equation}
The single-particle Hamiltonian $h(k)$ in momentum space
reads (setting the Fermion velocity as unit)
\begin{equation}
h(k)=\sum_{i=1,2}\mathbf{\alpha }_{i}\mathbf{k}_{i},\text{ }
\alpha _{1}=-\tau _{3}\sigma _{2},\text{ }\alpha _{2}=\tau _{3}\sigma _{1},
\end{equation}
where the $2\times 2$ Pauli matrices $\tau_{i}$ act on the valley
index, and $\sigma_{i}$ act on the sublattice index. Any
mass matrix $\mathcal{M}$ entering the single particle Hamiltonian obeys the relations
\begin{equation}
\left[ \alpha _{1,2},\mathcal{M}\right]_{+}=0, \quad \mathcal{M}^{2}=\mathbb{I}\Delta ,
\end{equation}
and the mass matrix $\mathcal{M}$ gapped the Fermionic spectral with the
form $E_k=\pm \sqrt{\mathbf{k}^{2}+\mathcal{M}^{2}}$. To describe more
electron ground states within the uniform framework of Dirac equation, we
can introduce another set of Pauli matrix $s_{i}$, which act on
the spin index. Taking spin, valley and sublattice
into account, the matrices $\mathbf{\alpha }_{i}(i=1,2)$ are $8\times 8$
Dirac matrices. The uniform descriptions of different order parameter with
gap generation will be constructed by defining the following
matrices,
\begin{eqnarray}
\alpha _{1} =s_{0}(-\tau _{3})\sigma _{2}, \quad  \alpha _{2} =s_{0}\tau _{3}\sigma _{1}.
\end{eqnarray}
Any mass matrix of the form $\mathcal{M}_{\mu \nu \lambda
}=$ $s_{\mu }\tau _{\nu }\sigma _{\lambda }$ must anticommutate with
matrices $\mathbf{\alpha }_{i}$, sixteen matrices match this property. In this paper, we define $\beta =s_{0}\tau _{0}\sigma _{3}$ as Dirac beta matrix, which realize mass-gap for quantum Hall effect rather than charge density wave as usual.

\subsection{Gap generation}

The microscopic symmetry may be spontaneously broke and lead to
semimetal-insulator transition in graphene-like Dirac semimetal. Thus, started from a four-fermions interaction,
the insulator can be builded from a nonvanishing fermion bilinear at mean field level. The nonzero fermion bilinear will lead to the mass-gap for the insulating phase, we consider the following continuum field theory
\begin{equation}
S(\bar{\psi}_{i},\psi _{i})=\int d^{3}x\left[ \bar{\psi}_{i}\gamma ^{\mu }i\partial _{\mu
}\psi _{i}+\frac{g}{2N}(\bar{\psi}_{i}\Gamma _{a}\psi _{j})^{2}\right] ,
\end{equation}
where $g$ labels the coupling strength for four-fermion interaction, and we have defined $\bar{\psi}_{i}\equiv \psi _{i}^{\dag }\gamma ^{0}$.
The three gamma matrices read
\begin{eqnarray}
\begin{split}
\gamma ^{0} =\beta =s_{0}\tau _{0}\sigma _{3}, \\
\gamma ^{1} =\beta \alpha _{1}=s_{0}\tau _{3}(i\sigma _{1}), \\
\gamma ^{2} =\beta \alpha _{2}=s_{0}\tau _{3}(i\sigma _{2}),
\end{split}
\end{eqnarray}
This Largrangian is similar to the Gross-Neven model with discrete chiral
symmetry that used to discuss the mass generation of the gauged vector
mesons in particle physics\cite{gn}. The gap generation or order profiles can
be studied by rewriting the functional integrals as
\begin{eqnarray}
Z &=&\int \mathcal{D}[\bar{\psi}_{i},\psi _{i},\phi ]e^{iS(\bar{\psi}
_{i},\psi _{i},\phi )},  \notag \\
S(\bar{\psi}_{i},\psi _{i},\phi ) &=&\int d^{3}x\left[ \bar{\psi}_{i}(\gamma
^{\mu }i\partial _{\mu }-\phi \Gamma _{a})\psi _{i}\mathcal{-}\frac{N\phi
^{2}}{2g}\right]. \label{mo}
\end{eqnarray}
The decoupling process of four-fermion interaction by introducing auxiliary boson
field in the functional integral is known as Hubbard-Stratonovich
transformation. Integrating out the fermion fields at large-$N$ limit, we get the effective potential for boson field as
\begin{equation}
V_{eff}(\phi )=\frac{N\phi ^{2}}{2g}+\frac{iN}{VT}\text{ln}\left[ \text{det}
(\gamma ^{\mu }i\partial _{\mu }-\phi \Gamma _{a})\right] .
\end{equation}
Eq.(\ref{mo}) show that $\phi\Gamma_a$ play the role of mass matrix and corresponds to an order pattern. In order to study the ground state behavior, we consider the static Higgs field.
Then, the effective potential is given by
\begin{eqnarray*}
V_{eff}(\phi ) &=&\frac{N\phi ^{2}}{2g}+\frac{iN}{(2\pi )^{3}}\int d^{3}k
\text{ln}(k^{2}-\phi ^{2})^{d/2}. \\
&=&\frac{N\phi ^{2}}{2g}-\frac{Nd}{4\pi ^{2}}\int_{c}^{\phi }d\phi
^{2}\left( \Lambda -\frac{\phi ^{2}}{2}\text{ln}\frac{\Lambda ^{2}+\phi ^{2}
}{\phi ^{2}}\right) ,
\end{eqnarray*}
where $d$ is the dimension of mass matrix and $\Lambda $ is an ultraviolet
cut-off in the regularization. Here we have used the relation $\text{det}(\gamma ^{\mu }i\partial _{\mu }-\phi \Gamma _{a})=(k^{2}-\phi^{2})^{d/2}$, which only determined by the dimension of the matrix.
By minimizing the potential $\partial
V_{eff}/\partial \phi =0$, the nonzero vacuum immediately gives the fermion mass
due to spontaneous symmetry breaking. The gap equation reads
\begin{equation}
\frac{1}{g(\Lambda )}=\frac{4}{\pi ^{2}}\int_{0}^{\Lambda }\frac{
k_{E}^{2}dk_{E}}{k_{E}^{2}+\phi ^{2}},
\end{equation}
from which we get the crtical coupling strength,
\begin{equation}
g_{c}=\pi ^{2}/4\Lambda .
\end{equation}
The mass-gap generate only if $g(\Lambda )>g_{c}$, and lead to an insulating phase.  To confirm the
fermion mass, setting $\phi =\phi (\Lambda ,g)$, we immediately find the
flow equation,
\begin{equation}
\frac{\partial g(\Lambda )}{\partial \Lambda }=-\frac{4g^{2}(\Lambda )}{\pi
^{2}}\left( 1-\frac{\Lambda \phi ^{2}}{\Lambda ^{2}+\phi ^{2}}\right) .
\end{equation}
Moreover, the mass-gap satisfies
\begin{equation}
\left( \Lambda \frac{\partial }{\partial \Lambda }+\beta (g)\frac{\partial }{
\partial g}\right) \phi =0,\text{ }\beta (g)=\Lambda \frac{\partial g}{
\partial \Lambda }.
\end{equation}
This means that the mass-gap is irrelevent to the ultraviolet cut-off $
\Lambda $ and the running coupling $g$, this should be, because the fermion
mass term must be a renormalization group invariant.

\subsection{Mass matrix and symmetry classification}

When the symmetry is spontaneously broken, the matrix $\phi \Gamma _{a}$ opens a fermion gap
at two Dirac points. As shown above, the kinetic part allow $16$ mass matrices, and each corresponds to an insulating phase. To find the physical explanation of these matrices, it's useful to define the following chiral symmetry time reversal symmetry.
For the Hamiltonian with lattice fermion residing on different sublattice,
flipping the sign of one sublattice(for example sublattice-b) will reverse
the overall sign of the Hamiltonian. The mapping of postive energy to
negative energy can be implemented by the operator $\Theta =s_{0}\tau
_{3}\sigma _{3},$ and this symmetry named sublattice symmetry(SLS) or chiral
symmetry. If a chiral order parameter is presented, the Hamiltonian
preserve SLS when satisfy
\begin{equation}
\Theta H(k)\Theta^{-1} =-H(k).
\end{equation}
The time reversal is defined as $Tc_{\uparrow +}=c_{\downarrow -}$, $Tc_{\downarrow +}=-c_{\uparrow -}$,
and which can be implemented by the matrix $\mathcal{T}=is_{2}\tau _{1}\sigma
_{1.}$ The Hamiltonian preserve time reverse symmetry(TRS) when satisfy%
\begin{equation}
\mathcal{T}H^{\ast }(-k)\mathcal{T}^{-1}=H(k).
\end{equation}
The more symmetry classification of mass matrix was presented in Ref.~\onlinecite{ryu}, and
one may to the reference for details.

The Kekul$\acute{e}$ dimerization corresponds to the fermion bilinear residing on the
opposite ends of nearest bond with a wave vector connecting two Dirac
points, which can be realized in terms of $\tau _{1}\sigma _{0}$ and $\tau
_{2}\sigma _{0}$. Such modulation preserve SLS, the associated mass matrix
named Kekul$\acute{e}$-VBS order pattern in analogy to the quantum
dimer model. Since there are four spin matrices(Three Pauli matrix and a
identity matrix), we have total eight Kekul$\acute{e}$-VBS mass matrices(see table-I). Any modulation with
lattice fermion sitting on the same sublattice does not preserve SLS, the
electron patter corresponds to chage density wave, spin density wave,
quantum Hall effect, and spin-orbital coupling driven quantum spin Hall
effect. The total 16 mass matrices and the corresponding physical explanation were
listed in table-I.

\begin{table}[t]
\caption{The list of total 16 mass matrices with $8\times 8$ canonical
representation, and each mass matric corresponds to a microscopic order
parameter. Here, $s_{1}$,$s_{2}$,$s_{3}$ denote the matrix $s_{x}$,$s_{y}$,$s_{z}$. $\circ $ and $\times $ denote preserves and breaks
symmetry, respectively }
\begin{tabular}{cccc}
\hline\hline
\quad Mass matrix & \qquad order parameter\qquad & \quad SLS\quad & \quad
TRS \quad \\ \hline
$s_{0}\tau_{1}\sigma_{0}$ & ReVBS & $\circ$ & $\circ$ \\
$s_{1}\tau_{1}\sigma_{0}$ & ReVBS$_{x}$ & $\circ$ & $\times$ \\
$s_{2}\tau_{1}\sigma_{0}$ & ReVBS$_{y}$ & $\circ$ & $\times$ \\
$s_{3}\tau_{1}\sigma_{0}$ & ReVBS$_{z}$ & $\circ$ & $\times$ \\
$s_{0}\tau_{2}\sigma_{0}$ & ImVBS & $\circ$ & $\circ$ \\
$s_{1}\tau_{2}\sigma_{0}$ & ImVBS${x}$ & $\circ$ & $\times$ \\
$s_{2}\tau_{2}\sigma_{0}$ & ImVBS$_{y}$ & $\circ$ & $\times$ \\
$s_{3}\tau_{2}\sigma_{0}$ & ImVBS$_{z}$ & $\circ$ & $\times$ \\
&  &  &  \\
$s_{0}\tau_{0}\sigma_{3}$ & QHE & $\times$ & $\times$ \\
$s_{1}\tau_{0}\sigma_{3}$ & QSHE$_{x}$ & $\times$ & $\circ$ \\
$s_{2}\tau_{0}\sigma_{3}$ & QSHE$_{y}$ & $\times$ & $\circ$ \\
$s_{3}\tau_{0}\sigma_{3}$ & QSHE$_{z}$ & $\times$ & $\circ$ \\
&  &  &  \\
$s_{0}\tau_{3}\sigma_{3}$ & CDW & $\times$ & $\circ$ \\
$s_{1}\tau_{3}\sigma_{3}$ & N$\acute{e}$el$_{x}$ & $\times$ & $\times$ \\
$s_{2}\tau_{3}\sigma_{3}$ & N$\acute{e}$el$_{y}$ & $\times$ & $\times$ \\
$s_{3}\tau_{3}\sigma_{3}$ & N$\acute{e}$el$_{z}$ & $\times$ & $\times$ \\
&  &  &  \\ \hline
\end{tabular}%
\end{table}

\section{Competition orders, WZW term and duality}

Near the two sides of the quantum critical
point(QCP), the two ordered phases break different symmetry and the continuous phase transition associate with two competing orders. Once
suppressing one of the orders, which necessarily leads to the emergence of
another order. The above features may embodies in the isotropic non-linear sigma model with a topological term (for
example Wess-Zumino-Witten term and mutual Chern-Smions term\cite{kou2,vish2}
). We will see that the crucial feature of the theory is the
existence of anticommutating mass terms and mutual duality, which may play key role in the
understand of Landau-forbidden quantum criticality.

In this section, we discuss the typical N$\acute{e}$el-Kekul$\acute{e}$ VBS quantum phase
transition. To describe the competition between the antiferromagnetic and
Kekul$\acute{e}$ VBS state, we include the five components N$\acute{e}$el-Kekul$\acute{e}$ VBS unit vector
$\vec{n}=\vec{\phi}/|\vec{\phi}|$, with the constraint $|\vec{n}|^{2}=1$,
\begin{equation}
\vec{\phi}\equiv (\vec{N},\text{ReVBS, ImVBS}).
\end{equation}%
The $O(3)\times U(1)$ vector(defined on $S^{4}$ surface) in (2+1)d spacetime
do not have topological nontrivial configurations for $\pi _{3}(S^{4})=0$,
but it support pertubative topological term due to non-triviality of the
homotopy group $\pi _{4}(S^{4})=Z$, which is known as WZW term. Considering
the action
\begin{equation}
S[\psi ,\bar{\psi},\vec{n}]=\int d^{3}x\bar{\psi}(i\gamma ^{\mu }\partial
_{\mu }-m\Upsilon ^{l}n_{l})\psi ,
\end{equation}
the non-linear sigma field theory and topological WZW terms can be restored
perturbatively by integrating out the gapped fermions in the presence of the
slowly varying $\vec{n}$. Follows the spirit Abanov and Wiegmann\cite{weigman}, the
effective action $S[\vec{n}]$ for N$\acute{e}$el-Kekul$\acute{e}$-VBS vector is obtained
as
\begin{eqnarray}
e^{iS[\vec{n}]} &=&\int D[\psi ,\bar{\psi}]e^{-S[\bar{\psi},\psi ,
\vec{n}]},  \notag \\
S[\vec{n}] &=&-i\text{tr}\ln (i\gamma ^{\mu }\partial _{\mu
}-m\Upsilon ^{l}n_{l}).
\end{eqnarray}
The $1/m$ expansion leads to the following remarkable action, $
S[\vec{n}]=S_{0}[n_{l}]+S_{WZW}^{E}[\vec{n}],$
\begin{equation}
S_{0}[n_{l}]=\frac{1}{g}\int d^{3}x(\partial _{\alpha }n_{l})^{2},\text{ }%
\frac{1}{g}\sim m.
\end{equation}
\begin{eqnarray}
S_{WZW}^{E}[\vec{n}] &=&\frac{4\pi i}{4!\text{Area}(S^{4})}\epsilon ^{\alpha
\beta \tau \rho }\epsilon ^{abcde}\int d^{2}xd\tau \int_{0}^{1}d\rho \times
\notag \\
&&(\partial _{\alpha }n_{a})(\partial _{\beta }n_{b})(\partial _{\tau
}n_{c})(\partial _{\rho }n_{d})n_{e}. \label{wzw}
\end{eqnarray}
The details of the derivation of Eq.(\ref{wzw}) are included in the appendix A.
The appearance of the WZW topological term is the consequence of the duality
between N$\acute{e}$el and Kekul$\acute{e}$-VBS order, implying that the topological objects of one
phase carries the quantum number of another phase(see follow for details).
From above derivations, the N$\acute{e}$el-Kekul$\acute{e}$-VBS QCP is obtained by incorporating
the WZW term to the $SO(5)$ nonlinear sigma model.

The N$\acute{e}$el order breaks the $SU(2)$ spin rotation to $U(1)$ symmetry, then the
$O(3)$ vector $\vec{n}=(n_{1},n_{2},n_{3})$ for N$\acute{e}$el order should be defined
on $SU(2)/U(1)=S^{2}$. The Kekul$\acute{e}$-VBS breaks $C_{6}$ lattice rotation or
lattice translational symmetry, if one consider the complex Kekul$\acute{e}$-VBS order
parameter, the discrete symmetry may be enlarged to a enhanced $U(1)$
symmetry. As discussed in superfluid phase transition or XY model, for the
symmetry unrelated phase transition, each ordered phase can be driven by the
condensation of the topological defect of the ordered phase. With the CP(1)
parametrization: $n_{a}=z^{\dag }\sigma _{a}z$, $z=(z_{1},z_{2})^{t}$ is a
complex two-component spinor with the constraint $|z_{1}|^{2}+|z_{2}|^{2}=1$
and $z_{1,2\text{ }}$are fractionalized \textquotedblleft
spinon\textquotedblright\ fields, we observe the Skyrmion charge in the N$\acute{e}$el
order as a total gauge flux
\begin{equation}
q_{s}=\frac{1}{2\pi }\int d^{2}x\epsilon ^{ij}\partial _{i}a_{j},\text{ }%
a_{i}=-\frac{i}{2}[z^{\dag }\partial _{i}z-\partial _{i}z^{\dag }z].  \label{ks}
\end{equation}
The internal gauged $U(1)$ symmetry appears as $a_{j}\rightarrow
a_{j}+\partial _{j}f$, so the Skyrmion number is conserved. The Skyrmion is
precisely mapped to a "magnetic" flux quantum. Since the Kekul$\acute{e}$-VBS order
parameter preserve $U(1)$, it support topological defect in the form of
vortex. In the dual description of vortex defect in the Kekul$\acute{e}$-VBS order,
the vorticity play the role of "electric" charge
\begin{equation}
q_{v}=\int_{c}dl\partial
_{i}e_{i}=\int_{c}dl\epsilon ^{ij}\partial _{i}\partial _{j}a_{0},
\end{equation}
The scalar field $a_{0}$ is phase field in this case, therefore the N$\acute{e}$el and
Kekul$\acute{e}$-VBS dual to each other due to electromagnetic duality.

Because $n_{a}$ preserve a $U(1)$ gauge degree of freedom after $SU(2)$ is
broken, the spinon field $z_{\alpha }$ will be coupled to the dynamical $%
U(1) $ gauge field $a_{\mu }$, we obtain the CP(1) model $S_{z}=\int
d^{2}xd\tau L_{z}$ ,that describes the deconfined QCP between N$\acute{e}$el and
Kekul$\acute{e}$-VBS order
\begin{eqnarray}
L_{z} &=&|(\partial _{\mu }-ia_{\mu })z_{\alpha }|^{2}+s_{z}|z_{\alpha
}|^{2}+r_{z}|z_{\alpha }|^{4}  \notag \\
&&+\varkappa (\epsilon ^{\mu \nu \lambda }\partial _{\nu }a_{\lambda })^{2}.
\end{eqnarray}%
The study of critical point can be aided by introducing a matter fields and
gauge field $b_{\mu }$ for vortex defect in Kekul$\acute{e}$-VBS order. Once the
matter fields is condensed, the gauge field acquires a mass term $b_{\mu
}^{2}$ via Higgs machanism. Then, we derive the Maxwell term in $L_{z}$ by
incorporating the mutual BF Chern-Simons(CS) term
\begin{equation}
L_{mcs}=i\sqrt{2\varkappa }\epsilon ^{\mu \nu \lambda }b_{\mu }\partial
_{\nu }a_{\lambda }\text{,}
\end{equation}%
after integrating out $b_{\mu }$. The mutual CS term is the concentrated
reflection of mutual duality between N$\acute{e}$el and Kekul$\acute{e}$-VBS order. Under the
gauge transformation $b_{\mu }\rightarrow b_{\mu }+\partial _{\mu }f^{b}$,
the current $j_{\mu }^{a}=i\sqrt{2\varkappa }\epsilon ^{\mu \nu \lambda
}\partial _{\nu }\partial _{\lambda }f^{b}$, and the time component as the
vortex when $f^{a}$ sweep across the vortex singularity in Kekul$\acute{e}$-VBS order.
Thus, the charge-vortex duality is captured by the mutual CS term, or WZW
term in terms of order vector.

Now, let us turn to describe the phase transition. As shown in Eq.(\ref{ks}), the Skyrmion is conserved, it correspond to configuration of $a_{\mu }$
at which creats a $2\pi $ flux. Once the quantum flux is condensed, the
extra term induced by mutual CS is non-vanished, i.e., $\int d^{2}xd\tau
\epsilon ^{\mu \nu \lambda }\partial _{\mu }f^{b}\langle \partial _{\nu
}a_{\lambda }\rangle \neq 0$. The flux condensation destroys the Skyrmion
number, meanwhile, it spontaneously breaks the $U(1)_{b}$ symmetry and leads
to the Kekul$\acute{e}$-VBS order. The condensation of vortex in Kekul$\acute{e}$-VBS order
gives $\langle b_{\mu }\rangle \neq 0$, such condensation is irrelevent to
the gauge condition: $\int d^{2}xd\tau \epsilon ^{\mu \nu \lambda }\partial
_{\nu }\partial _{\lambda }f^{a}\langle b_{\mu }\rangle =0$, it does not
destroy conservation law
\begin{equation}
\partial_{\mu} j^{\mu}_a=0.
\end{equation}
The spinon fields $z_{\alpha }$ preserve the $U(1)_{a}$ symmetry and whose
condensation defined on $SU(2)/U(1)=S^{2}$, which is equivalent to N$\acute{e}$el
order.

As discussed by Xu in Ref.~\onlinecite{xu}, the Skyrmion of N$\acute{e}$el order carries
the quantum number of Kekul$\acute{e}$-VBS order, the condensate of Skyrmion breaks
N$\acute{e}$el order and it also induces Kekul$\acute{e}$-VBS order. To make the N$\acute{e}$el-Kekul$\acute{e}$-VBS
duality explicitly, we parametrize the order vector $\vec{n}$ as
\begin{equation}
\vec{n}=[\phi (r)\vec{v}\mathbf{,}\sqrt{1-\phi ^{2}(r)}\vec{s}(\tau ,\rho )],
\end{equation}
with $\phi (0)=0$ and $\phi (\infty )=1$. Integrating over the space lead to
the effective action near the core of vortex defect,
\begin{equation}
S^{\text{v}}=\frac{i}{4}\int d\tau d\rho \epsilon ^{\alpha \beta
}\epsilon ^{abc}s_{a}(\partial _{\alpha }s_{b})(\partial _{\beta }s_{c}),
\end{equation}
which is right the WZW term for a $1/2$-spin in (0+1)D, that is $1/2$%
-spinon. Therefore, the vortex defect carries the $1/2$-spinon quantum
number. It's known that the Kekul$\acute{e}$-VBS order enjoy the spectral reflection,
symmetry and the vortex defect host single fermionic midgap zero energy mode
in the core. If the system preserve the time-reversal symmetry, the Kramers
conjugation ensure that there are two zero energy levels, each carries up $
1/2$-spinon quantum number and down $1/2$-spinon quantum number,
repectively. The occupation of these zero modes leads to fractionalization
of electrons, which gives the following four types of states:
\begin{eqnarray}
\begin{split}
f(+1/2 \uparrow,+1/2\downarrow )&(\Delta Q=1,S_{z}=0), \\
f(-1/2 \uparrow,-1/2\downarrow )&(\Delta Q=-1,S_{z}=0), \\
f(+1/2 \uparrow,-1/2\downarrow )&(\Delta Q=0,S_{z}=1/2), \\
f(-1/2 \uparrow,+1/2\downarrow )&(\Delta Q=0,S_{z}=-1/2).
\end{split}
\end{eqnarray}
The states are right the spin-charge separation holon, chargeon, and the two
spinon states studied in quantum spin Hall effect\cite{ran,qi1}, and all the
viewed particles are bosons. The self duality also
result to spin-charge separation. Moreover, the condenses of these bosons
will help us understand the rich phase transition like spin liquid-magnetic,
superconductor-magnetic transition, etc.

By using sign-free Majorana quantum Monte Carlo simulations, Ref.~\onlinecite{yao}
 show the evidences of deconfined QCP between antiferromagnetic and
Kekul$\acute{e}$-VBS transition. Based on present understands, the WZW term or mutual
CS term guarantee the duality between N$\acute{e}$el and Kekul$\acute{e}$-VBS order, and the
phase transition is described by the fractionalized fields $z_{\alpha }$.
Therefore, we conclude that the QCP between N$\acute{e}$el-Kekul$\acute{e}$-VBS transition is
\textit{deconfined}, and such QCP may be generized to other
multiple-components order parameter.

\section{The index theorem and Neel-Kekule-VBS duality}

The WZW term exhibits mutual duality between N$\acute{e}$el and Kekul$\acute{e}$-VBS order, we now turn our attention to how the duality relate to index of the topological excitation. For the system with chiral symmetry, there exist a index theorem relate the topological stability to the number of the zero modes hosting by the topological excitation.
Here we first briefly review the Jackiw-Rossi-like mode and the index theorem. And beyond any symmetry, the topological index for twisted order is discussed in the sense of N$\acute{e}$el-Kekul$\acute{e}$-VBS duality.

\subsection{The nontrivial twisted order pattern}

The Kekul$\acute{e}$-VBS order pattern corresponds to the modulation of fermion bilinear operator with
different sublattice and Dirac point, and thus preserve chiral symmetry.
Without considering the spin degrees of freedom, if $\psi $ enter as the Kekul$\acute{e}$-VBS
mass matrix, the matrix fields V$_{ab}=\psi ^{\dagger }\mathcal{M}_{ab}\psi $
condenses, for instance
\begin{equation}
\left\langle V_{10}\right\rangle =|\Delta
|\left\langle \psi ^{t}\tau _{1}\sigma _{0}\psi \right\rangle \neq 0,
\end{equation}
such condenses gapped out the nodal quasiparticles. In order to demonstrate the
connection between the twist Kekul$\acute{e}$-VBS order pattern and midgap zero energy states
at the Dirac equation level, it's useful to suppose that the order pattern could
be slowly varying on the scale of the lattice spacing, and the nontrivial
background topology will be realized by the complex valued mass matrix
field. In continuum limit, the complex Kekul$\acute{e}$-VBS
mass matrix as $V=$Re$V$+$i$Im$V$, with Re$V$$=\Delta \left\langle \psi
^{t}\tau _{1}\sigma _{0}\psi \right\rangle $ and Im$V$$=\Delta \left\langle \psi
^{t}\tau _{2}\sigma _{0}\psi \right\rangle $. Thus, the
fermion interacting with the twist Kekul$\acute{e}$-VBS order according to the
Lagrangian
\begin{eqnarray}
L_{\text{VBS}} &=&\bar{\psi}\gamma ^{\mu }(i\partial _{\mu }+\gamma
_{5}a_{\mu })\psi -\bar{\psi}\beta V\psi,  \notag \\
V&=&\beta _{1}\text{cos}\theta +\beta _{2}\text{sin}\theta,
\end{eqnarray}
where $\beta=\tau_0\sigma_3,,\beta_1=\tau_1\sigma_0, \beta_2=\tau_2\sigma_0$, and the gauge potential we choose with vanishing temporal component $a_0=0$.
The phase twist in such Kekul$\acute{e}$-VBS mass matrix realize domain wall or vortex at two
dimensional spatial, indeed, the background topology for the Lagrangian equal to the Jackiw-Rossi model
that describe charged fermion interacting with the scalar fields of the two
dimensional Abelian Higgs model.

It's known that the soliton excitations are associated with many profound physical phenomena like Fractionalization, topological degeneracy and zero energy quasiparticle.
The occupation of zero energy soliton lead to spin-charge separation, before discussing the wavefunction of the zero mode solutions, we first make
some discussion about the Lagrangian $L_{\text{VBS}}$ and the Fu-Kane model that describe the surface of topological insulator with
a conventional superconductor proximity to it in terms of the Hamiltonian
\begin{equation}
H_{\text{FK}}=\tau_z\mathbb{I}(\sigma _{i}k_{i}-\mu )+\Delta(\beta _{1}\text{cos}\theta +\beta _{2}\text{sin}\theta ).
\end{equation}
$H_{\text{FK}}$ does not breaking time reversal symmetry, and the Pauli
matrix $\tau_z\mathbb{I}$ mixed the particle and hole part.
When the chemical potential satisfy $\mu >|\Delta|,$ due to the $\mathbb{Z}_2$ topological index, there is at least
one topologically protected zero mode localized on the odd vorticity vortex.
When $\mu =0,$ the massive vortex (non-trivial twist of background field)
also guarantee the existence of zero modes and now the model equal to $L_{\text{VBS}}$.

Many literatures have studied the single-valued and normalizable wave
functions for zero energy solution in the presence of nontrivial twist fields. For $|n|$-twisted order
parameter, there are $|n|$ independent normalizable zero energy states, the property and the special form for each zero energy solution may differ.
The Hamiltonian subjected to the Kekul$\acute{e}$-VBS order pattern is given by
\begin{equation}
H=\alpha _{i}(-i\partial _{i}-\gamma _{5}a_{i})+\Delta(r)(\beta _{1}\text{
cos}\theta +\beta _{2}\text{sin}\theta ), \label{vbh}
\end{equation}
and the normalizable zero energy solutions have been discussed in appendix B.
When the degree of twisting for the Kekul$\acute{e}$-VBS order equal to $2k$, there are $2k$
normalizable zero modes, and each zero energy mode spinors characterized by two phase dependence $\psi_{\mathbf{k}a}=fe^{il_1\theta}+ge^{il_2\theta}$
($l_{1}\neq l_{2}$ or $m_{1}\neq m_{2}$); When the whole twist of the Kekul$\acute{e}$-VBS order equal to $2k+1$, there are $2k+1$ zero modes, among which only one for single phase dependence($l_{1}=l_{2})$ and the others characterized by two phase dependence. The
normalizable single phase dependence wave function for zero mode on sublattice-a can be easily obtained as
\begin{equation}
\begin{split}
\psi_{\mathbf{k}a}&=\mathcal{N}_{a}e^{-\int_{0}^{r}(\Delta (r)+\frac{2na(r)-(n-1)}{2r})dr}
e^{i[\frac{(n-1)\theta }{2}-\frac{\pi }{4}]},\\
\psi_{\mathbf{k'}a}&=i\bar{\psi}_{\mathbf{k}a}.
\end{split}
\end{equation}
Since $\Delta (r)$ and $a(r)$ vanish at small $r$, the wave function to be normalizable
require $n\leq -1$. Similarly, the normalizable single-phase dependence zero mode on sublattice-b as
\begin{equation}
\begin{split}
\psi_{\mathbf{k}b}&=\mathcal{N}_{b}e^{-\int_{0}^{r}(\Delta (r)+\frac{(n+1)-2na(r)}{2r})dr}
e^{i[\frac{(n+1)\theta }{2}+\frac{\pi }{4}]},\\
\psi_{\mathbf{k'}b}&=-i\bar{\psi}_{\mathbf{k}b}.
\end{split}
\end{equation}
The wave function at small $r$ to be normalizable require $n\geq 1$, and same condition can be seen from Ref.~\onlinecite{mudry}. Since the sublattice symmetry (or named chiral symmetry), the zero mode wave function on sublattice-a also hold on sublattice-b with the substitutions $\psi_{\mathbf{k}a}\rightarrow \psi_{\mathbf{k}b}$ and $n \rightarrow -n$.
The chiral operation is defined as $\Gamma_{5}=\tau _{3}\sigma _{3}$, which gives $\Gamma _{5}\psi _{a/b}=\pm1$. The sublattice symmetry ensure that the energy eigenstate $\pm E$ come into pairs, thus the zero modes bound to the VBS vortex are classified by $\mathbb{Z}_2$. Only for odd twisting, there is at least
a topologically protected Fermion zero mode. If the system possess
particle-hole symmetry, which will lead to Majorana zero mode, such as topological superconductor studied in
Ref.~\onlinecite{cheng}.

\subsection{The topological index and the N$\acute{e}$el-Kekul$\acute{e}$ VBS duality}
The index theorem bridge the total number and the topological stability of the zero modes, which relate the analytical index of a Dirac operator to the the
winding number of background scalar field in 2D spatial. It's easily
show that the Hamiltonian in Eq.(\ref{vbh}) connect with an elliptic
differential operator for vanishing energy. For the system
with chiral symmetry, the mid-gap states are always come into pairs, and the
difference between the number of zero mode with opposite chirality give the
analytical index $\text{ind}H=n_+-n_-$. The eigen function for vanishing energy is
\begin{equation*}
\sigma _{2}(i\partial _{1}+a_{1})\psi -\sigma _{1}(i\partial _{2}+a_{2})\psi
+\Delta \sigma _{2}\psi ^{\ast }=0.
\end{equation*}
By writing $\psi (x)$ as two real function $\psi(x)=\psi_1(x)+i\psi_2(x)$, and expressing $\Delta=\Delta_1+i\Delta_2$,  the corresponding differential operator is of the form
\begin{equation*}
\mathcal{D}=(-\partial _{2}-i\tau _{2}\partial _{1})+(a_{1}-i\tau
_{2}a_{2})+(\tau _{3}\Delta _{1}+\tau _{1}\Delta _{2})
\end{equation*}
The index theorem for $\mathcal{D}$ gives the analytical index\cite{weinberg}.

The index theorem is applicable only when the system preserve chiral
symmetry. Beyond chiral symmetry, the N$\acute{e}$el-Kekul$\acute{e}$ VBS duality provide us an alternative way to derive topological index via dual topological excitation.
For present (2+1)D Dirac fermion interacting with the background scalar fields(complex Kekul$\acute{e}$-VBS order), the dual target space can be obtained by considering the extended order parameter $\phi =($
Re$\Delta,$Im$\Delta,h)$ according to the Lagrangian
\begin{equation}
L=\bar{\psi}i\gamma ^{\mu }(\partial _{\mu }-i\gamma _{5}a_{\mu })\psi -\bar{%
\psi}\left( \sum_{k=1}^{3}\Gamma _{k}\phi _{k}\right) \psi .
\end{equation}
As shown in appendix C, the topological current
\begin{eqnarray}
J^{\mu } =-\frac{1}{8\pi \phi ^{3}}\epsilon ^{\mu \nu \lambda }\epsilon ^{abc}\phi
_{a}\partial _{\nu }\phi _{b}\partial _{\lambda }\phi _{c},
\end{eqnarray}
and the topological number $Q=\int d^{2}xJ^{0}$ classifies all the
homotopy classes $\pi _{2}(S^{2})$, which is equivalent to the degrees of
mapping from $T^{2}$ to $S^{2}$. We suppose the VBS
vortex $\phi_v =($Re$\Delta, $Im$\Delta)$ lie on the plane where $h=0$.
The half-degrees of mapping from $T^{2}$ to $S^{2}$($h>0$)can be
separated into two parts,

\begin{equation}
\frac{\Omega }{2}=\int d^{2}xJ^{0}(a=3)+\int d^{2}xJ^{0}(a\neq 3).
\end{equation}%
One can easily show that(by direct computations), the second term vanishes
when $h$ is regularized to zero. Now the half-degrees of mapping then becomes
\begin{eqnarray}
\frac{\Omega }{2} &=&-\int d^{2}x\epsilon ^{ij}\epsilon ^{ab}h\partial
_{i}\phi _{a}\partial _{j}\phi _{b}/8\pi \phi ^{3}  \notag \\
&=&-\int h\epsilon ^{ab}d\phi _{a}\wedge d\phi _{b}/4\pi \phi ^{3}
\end{eqnarray}
Writting $\phi =\Delta (r)\hat{\phi}(\theta )$, so $d\phi _{a}\wedge d\phi
_{b}=\partial _{i}\phi _{a}\partial _{j}\phi _{b}dr\wedge d\theta $, now the
half-degrees of mapping is given by
\begin{eqnarray}
\frac{\Omega }{2} &=&-\frac{1}{4\pi }\int_{0}^{\infty }\frac{
hd\Delta^2 }{(\Delta ^{2}+h^{2})^{3/2}}\int d\theta \epsilon
^{ab}\epsilon ^{ab}\hat{\phi}_{a}\partial _{\theta }\hat{\phi}_{b}  \notag \\
&=&\frac{1}{4\pi }\int d\theta \epsilon ^{ab}\epsilon ^{ab}\hat{\phi}
_{a}\partial _{\theta }\hat{\phi}_{b}
\end{eqnarray}
Thus we obtain the well known result
\begin{equation}
\Omega =\frac{1}{2\pi }\int d\theta \epsilon ^{ab}\epsilon ^{ab}\hat{\phi}
_{a}\partial _{\theta }\hat{\phi}_{b},
\end{equation}
this quantum number gives the difference between the number of zero mode with
opposite chirality, from which the number of zero modes can be identified.
Thus, the analytic index of a Dirac operator relate to a topological index of an associate mapping
in its dual topological defect, at least valid for present case. The N$\acute{e}$el-Kekul$\acute{e}$-VBS duality imply that there exist an invariant connects the following two theories:
\begin{eqnarray}
L_1&=&\bar{\psi}i\gamma\cdot(\partial-i\gamma_5a)\psi+m(r)e^{i\gamma_5\theta}\bar{\psi}\psi+\cdots,\\
L_2&=&\bar{\psi}i\gamma\cdot(\partial-ib)\psi+\bar{\psi}\vec{n}\cdot\vec{\sigma}\psi+\cdots.
\end{eqnarray}

\section{Conclusions and discussion}

In conclusion, motivated by the predictable conclusion that the quantum
critical point between Kekul$\acute{e}$ valence-bond-solid(VBS) and antiferromagnetic
is deconfined in Ref.~\onlinecite{yao}, we have studied the dynamical spontaneous
symmetry breaking in two dimensional graphene Dirac semimetal, and discussed the N$\acute{e}$el-Kekul$\acute{e}$ VBS mutual duality and the quantum transition
between them. We consider the general quartic fermion-bilinear
interactions that will generate rich order parameters when the dynamical
symmetry is broken. One feature in our study is the appearance of multicritical
point for various orders, which is shown to irrelevant to the special form of
quartic fermion-bilinear but only depends on the dimensions of the mass
matrix(each mass matrix corresponds to an order parameter). The multicritical point exhibits the Landau-forbidden quantum phase
transition, and the key ingredient is the anticommutativity among these mass matrices.
Very recently, we observed that Ref.~\onlinecite{ass} have numerically shown the multicritical and the continuous
N$\acute{e}$el-Kekul$\acute{e}$ VBS transition, which is consistent with our results.

Further, we have shown that the typical N$\acute{e}$el-Kekul$\acute{e}$ VBS quantum criticality can be
understood in term of the mutual duality between them. By using the
'superspin' non-linear sigma field theory, the N$\acute{e}$el-Kekul$\acute{e}$ VBS quantum
criticality is captured by the WZW action or self BF Chern-Simons field
theory, which reveal the mutual duality for N$\acute{e}$el and Kekul$\acute{e}$ VBS order.
Statically speaking, the mutual duality in the sense that the topological
defect in either phase carries the quantum number of another phase.  Dynamically, the mutual duality in the sense that the order
parameter in either phases carries the symmetry of another phase, which
embodies in the Chern-Simons action. The breaking of symmetry in one
phase meanwhile breaks the dual symmetry and lead to the emergence of
another order. Since the transition is described by the CP(1) fields coupled
with $U(1)$ gauge field, the quantum critical point for N$\acute{e}$el-Kekul$\acute{e}$
transition is deconfined. We also saw that the mutual duality
lead to the profound phenomenon of spin-charge separation, and the concept of duality may help
us understand more rich physics.

We now make some comments about the concept of duality. The N$\acute{e}$el-Kekul$\acute{e}$ VBS duality in present paper means self-duality near two sides of the critical point rather than duality between two seemingly different theories. Various recent study concentrate on the duality between the theoretical fields theory in high-energy physics and the quantum criticality in condensate physics. For example, the $N_f=2$ noncompact QED$_3$ and the easy-plane $U(1)$ noncompact CP$^1$ model\cite{mzy,cwang}(describing the N$\acute{e}$el-VBS transition in magnets with XY spin symmetry), the QED$_3$-Gross-Neveu model and $SU(2)$ noncompact CP$^1$ model\cite{hyc}. The duality in the sense that both two theories are described by the same effective low-energy
field theory and are characterized by same critical behavior at criticality. The hidden duality relate the interacting driven topological quantum phase transition and the quantum criticality,
and which will deepen our consensus for the fundament.

Finally, let us give a summary and prospect for Landau-forbidden quantum
criticality, which features the follows: (i) Two phases are symmetry-incompatible. (ii) There exist mutual duality between the two phases,
as stated in the above. (iii) The space-time topological
excitations must be involved in the phase transition process. (iv) The phase
transition from phase-B to phase-A may destroy some structure(like entanglement or interacting symmetry protected topological phase)
belong to the phase-B. The study of FIQC\cite{yao} and symmetric mass generation(many-body gap\cite{yyz}) give some new insights for us, and the duality might shed some light on this topic. More, the surface realization of
Landau-beyond quantum
criticality for strong correlated topological
states in high dimensional is also an interesting issue to be explored.

\begin{acknowledgments}
We thank Martin Hohenadler for helpful email communications. This work is supported by NSFC under the Grant No.11647111(J.Zhou), No.11504285(Y.J.Wu), and No.11474025, 11674026(S.P. Kou). We are also
grateful to the supports from the research start-up funds of Guizhou University with No. 201538(J. Zhou).
\end{acknowledgments}

\begin{appendix}
\section{Derivation of nonlinear sigma model with WZW term}
This appendix is aim to derive the low energy effective theory for the N$\acute{e}$el-Kekul$\acute{e}$ VBS SO(5) order parameters. The perturbation method which used in the derivation named gradient expansion or large-$m$ expansion, and such procedure had been extensively adopted. In the presence of slowly varying background, the total action included Fermi and Boson fields reads
\begin{equation}
S[\psi ,\bar{\psi},\vec{n}]=\int d^{3}x\bar{\psi}(i\gamma ^{\mu }\partial
_{\mu }-m\Upsilon ^{l}n_{l})\psi ,
\end{equation}
where $\Upsilon^{1}=s_1\tau_3I$, $\Upsilon^{2}=s_2\tau_3I$, $\Upsilon^{3}=s_3\tau_3I$, $\Upsilon^{4}=I\tau_1\sigma_3$, $\Upsilon^{5}=\tau_2\sigma_3I$ and $I$ denotes the $2\times2$ identity matrix.

Following the spirit of Abanov and Wiegmann\cite{weigman}, the superspin nonlinear sigma model and the WZW term can be derived by integrating over the fermions. When gapped fermions are integrated out, the effective action for the boson fields is obtained as
\begin{eqnarray}
e^{iS_{e}[\vec{n}]} &=&\int D[\psi ,\bar{\psi}]e^{-S[\bar{\psi},\psi ,
\vec{n}]}  \notag \\
S_{e}[\vec{n}] &=&-i\text{tr}\ln (i\gamma ^{\mu }\partial _{\mu
}-m\Upsilon ^{l}n_{l}).
\end{eqnarray}
The variation $\delta S_{e}[\vec{n}]$ respect to $n_{l}$ gives
\begin{eqnarray*}
\delta S_{e}^{(n)}[\vec{n}] &=&-i\text{tr}(i\gamma ^{\alpha }\partial
_{\alpha }-m\Upsilon ^{a}n_{a})^{-1}(-m\Upsilon ^{l}\delta n_{l}) \\
&=&-i\sum_{n}(-1)^{n}\text{tr}\{[(\partial ^{2}+m^{2})^{-1}(-im\Upsilon
^{c}\gamma ^{\alpha }\partial _{\alpha }n_{c})]^{n} \\
&&\times (\partial ^{2}+m^{2})^{-1}(-i\gamma ^{\beta }\partial _{\beta
}-m\Upsilon ^{s}n_{s})(-m\Upsilon ^{l}\delta n_{l})\},
\end{eqnarray*}
where the trace represents the trace over the spatial and momenta coordinates. It's easily observed that the $\delta S_{e}^{(1)}[\vec{n}]$ is
non-vanished,
\begin{eqnarray}
\delta S_{e}^{(1)}[\vec{n}] &=&im^{2}\text{tr[}1_{8}\text{]tr}
_{k}[(m^{2}-k^{2})^{-2}\partial _{\alpha }n_{l}\partial _{\alpha }\delta
n_{l}] \notag \\
&=&\frac{m}{2\pi }\int d^{3}x\partial _{\alpha }n_{l}\delta (\partial
_{\alpha }n_{l}),
\end{eqnarray}
$k^{2}=k_{0}^{2}-\vec{k}^{2}$, $\delta S_{e}^{(1)}[\vec{n}]$ generate the SO(5)
nonlinear sigma model for vector $\hat{n}$ that combines the N$\acute{e}$el-Kekul$\acute{e}$ VBS
order parameters,
\begin{equation}
S_{e}^{(1)}[\vec{n}]=\frac{1}{g}\int d^{3}x(\partial _{\alpha }n_{l})^{2},
\frac{1}{g}\sim m.
\end{equation}
The leading-order correction to the low-energy effective action is $\delta
S_{e}^{(3)}[\vec{n}]$, which is computed as
\begin{eqnarray}
\delta S_{e}^{(3)}[\vec{n}] &=&\mathcal{N}\text{tr(}\Upsilon ^{a}\gamma
^{\alpha }\Upsilon ^{b}\gamma ^{\beta }\Upsilon ^{c}\gamma ^{\gamma
}\Upsilon ^{d}\Upsilon ^{e}\text{)}\times  \notag \\
&&\int d^{3}x(\partial _{\alpha }n_{a})(\partial _{\beta }n_{b})(\partial
_{\sigma }n_{c})n_{d}(\delta n_{e}), \label{A5}
\end{eqnarray}
where the integrand
\begin{eqnarray}
\mathcal{N}=\int \frac{d^{3}k}{(2\pi )^{3}}\frac{m^{5}}{
(m^{2}-k^{2})^{4}}=\frac{2\pi i}{4!\text{Area}(S^{4})}.
\end{eqnarray}
The trace in the right of Eq.(\ref{A5}) indeed a product of two anti-symmetrical tensors, which gives by
\begin{equation}
\text{tr}(\Upsilon ^{a}\gamma ^{\alpha }\Upsilon ^{b}\gamma ^{\beta
}\Upsilon ^{c}\gamma ^{\tau }\Upsilon ^{d}\Upsilon ^{e})=8i\epsilon ^{\alpha
\beta \tau }\epsilon ^{abcde}.
\end{equation}

Let $\vec{n}(x,\rho )$ distort continuously such that $\vec{n}(x,0)=(n_{1}=1,%
\vec{0})$ and $\vec{n}(x,\rho =1)=\vec{n}(x)$, the $\vec{n}(x)$ is thus
defined on the four dimentional spacetime disk $D^{4}$ with boundary $%
\partial D^{4}=S^{3}$. Therefore, the field $\vec{n}(x)$ defines a map from $%
S^{3}$ to $S^{4}$. We can introduce an auxiliary variable $\rho \in \lbrack 0,1]$,
then $\delta S_{e}^{(3)}[\vec{\phi}]$ is rewritten as
\begin{eqnarray}
\delta S_{e}^{(3)}[\vec{n}] &=&8i\mathcal{N}\epsilon ^{\alpha \beta
\sigma }\epsilon ^{abcde}\int d^{3}x\int_{0}^{1}d\rho \times  \notag \\
&&\partial _{\rho }[(\partial _{\alpha }n_{a})(\partial _{\beta
}n_{b})(\partial _{\tau }n_{c})n_{d}(\delta n_{e})],
\end{eqnarray}
which indeed the functional derivative of the WZW term. The fields variable
only defined on the boundary, note that the term
\begin{equation}
\epsilon ^{\alpha \beta
\sigma }\epsilon ^{abcde}\int d^{3}x\int_{0}^{1}d\rho\times\partial _{\alpha }n_{a}\partial _{\beta
}n_{b}\partial _{\tau }n_{c}\partial _{\rho }n_{d}\delta n_{e},
\end{equation}
need to vanish, since for the small changes, $\vec{n}\rightarrow\vec{n}+\delta\vec{n}$, $\delta n_{e}$ is perpendicular to $n_{e}$. Combined all these results lead to the Euclidean WZW action $S_{\text{WZW}}^E[\vec{n}]=2\pi i\mathcal{S}$,
\begin{eqnarray}
\mathcal{S} &=&\frac{2}{4!\text{Area}(S^{4})}\epsilon ^{\alpha \beta \tau \rho
}\epsilon ^{abcde}\int d^{2}xd\tau \int_{0}^{1}d\rho \times  \notag \\
&&(\partial _{\alpha }n_{a})(\partial _{\beta }n_{b})(\partial _{\tau
}n_{c})(\partial _{\rho }n_{d})n_{e}.
\end{eqnarray}
\section{Derivation of the zero mode under twist order}
In this appendix, we discuss the zero-energy modes for fermion-vortex system with $n$-twist background fields. The Hamiltonian subjected to a complex Kekul$\acute{e}$ VBS order parameter(VBS order pattern with a real and imaginary part) is given by
\begin{equation}
H=\alpha _{i}(-i\partial _{i}-\gamma _{5}a_{i})+\Delta(\beta _{1}\text{
cos}\theta +\beta _{2}\text{sin}\theta ),
\end{equation}
where $\alpha _{1}=-\tau _{3}\sigma _{2}$, $\alpha _{2}=\tau _{3}\sigma _{1}$, and the eigenvalue equation reads $H\psi _{0}=0$ for zero energy modes, with
\begin{equation}
\psi _{0}=(\psi _{1},\sigma _{2}\psi _{1}^{\ast })^{t}.
\end{equation}
We assume the system is isotropic and set $\psi _{1}=[f(r,\theta ),g(r,\theta )]^{t
}$, then the vanishing energy solutions is determined by
\begin{equation}
\begin{split}
\lbrack \partial _{+}+(a_{2}-ia_{1})]f-i\Delta f^{\ast }& =0, \\
\lbrack \partial _{-}-(a_{2}+ia_{1})]g+i\Delta g^{\ast }& =0,
\end{split}
\end{equation}%
where $\partial _{\pm }=e^{\pm i\theta }(\partial _{r}\pm i\partial _{\theta
}/2)$. We consider a $n$-twists of the scalar field, so the decomposition of
$\Delta $ in polar coordinates is $\Delta =\Delta (r)e^{in\theta },$ $\Delta
(r)$ vanishs at the origin and tend to a constant at infinity. The static
axial gauge field $\vec{a}$ integral over a closed loop that encircles the
origin yields the degree of twisting, which we can parameterize as
\begin{equation}
a_{j}=-n\varepsilon _{ij}r_{j}a(r)/r^{2},a(r\rightarrow \infty )=1/2.
\end{equation}
The existence of gauge field only play the role of changing the profile of
background but not the zero energy eigenvalue. Using the polar coordinates,
the eigen equation for vanishing energy take the form
\begin{equation}
\begin{split}
e^{i\theta }[\partial _{r}+\frac{i\partial _{\theta }}{r}+\frac{na(r)}{r}
]f-i\Delta f^{\ast }& =0, \\
e^{-i\theta }[\partial _{r}-\frac{i\partial _{\theta }}{r}-\frac{na(r)}{r}
]g+i\Delta g^{\ast }& =0.
\end{split}
\end{equation}
The general ansatz for the zero energy eigenvectors may be chosen as
\begin{equation}
\begin{split}
f(r,\theta )& =e^{-i\frac{\pi }{4}}[f_{1}(r)e^{il_{1}\theta
}+f_{2}(r)e^{il_{2}\theta }], \\
g(r,\theta )& =e^{i\frac{\pi }{4}}[g_{1}(r)e^{im_{1}\theta
}+g_{2}(r)e^{im_{2}\theta }],
\end{split}
\end{equation}
where both$f_{i}(r)$ and $g_{i}(r)$ are real. The radial equations for $f_{i}(r)$ then becomes
\begin{equation}
\begin{split}
(\partial _{r}-\frac{l_{1}}{r}+\frac{na(r)}{r})f_{1}+\Delta (r)f_{2}& =0, \\
(\partial _{r}-\frac{l_{2}}{r}+\frac{na(r)}{r})f_{2}+\Delta (r)f_{1}& =0,
\end{split}
\end{equation}
the corresponding compatibility condition satisfies $l_{1}+l_{2}=n-1$.
And the radial equations for $g_{i}(r)$ becomes
\begin{equation}
\begin{split}
(\partial _{r}+\frac{m_{1}}{r}-\frac{na(r)}{r})g_{1}+\Delta (r)g_{2}& =0, \\
(\partial _{r}+\frac{m_{2}}{r}-\frac{na(r)}{r})g_{2}+\Delta (r)g_{1}& =0,
\end{split}
\end{equation}
the corresponding compatibility condition is $m_{1}+m_{2}=n+1$.

The index theorem state that the analytical index is identical to the the number of zero modes, and also identical to the winding number of the
background fields. Here the normalizable zero mode is given by $\psi_0=(\psi_{\mathbf{k}a},\psi_{\mathbf{k}b},\psi_{\mathbf{k'}b},\psi_{\mathbf{k'}a})^t$. According to the compatibility condition, if the whole twist of the Kekul$\acute{e}$-VBS order is even, all the normalizable zero modes are characterized by two-phase dependence($l_{1}\neq l_{2}$ or $m_{1}\neq m_{2}$); If the
whole twist of the order is odd, one of normalizable
zero modes is characterized by single-phase dependence($l_{1}=l_{2}$),
and the others $n-1$ normalizable modes are two-phase dependence.
The single phase dependence wave function for zero mode on sublattice-a is
\begin{equation}
\begin{split}
\psi_{\mathbf{k}a}&=\mathcal{N}_{a}e^{-\int_{0}^{r}(\Delta (r)+\frac{2na(r)-(n-1)}{2r})dr}
e^{i[\frac{(n-1)\theta }{2}-\frac{\pi }{4}]},\\
\psi_{\mathbf{k'}a}&=i\bar{\psi}_{\mathbf{k}a}.
\end{split}
\end{equation}
Since $\Delta (r)$ and $a(r)$ vanish at small $r$, the wave function to be normalizable
require $n\leq -1$. Similarly, the normalizable single-phase dependence zero mode on sublattice-b as
\begin{equation}
\begin{split}
\psi_{\mathbf{k}b}&=\mathcal{N}_{b}e^{-\int_{0}^{r}(\Delta (r)+\frac{(n+1)-2na(r)}{2r})dr}
e^{i[\frac{(n+1)\theta }{2}+\frac{\pi }{4}]},\\
\psi_{\mathbf{k'}b}&=-i\bar{\psi}_{\mathbf{k}b}.
\end{split}
\end{equation}
The wave function at small $r$ to be normalizable require $n\geq 1$. With the substitutions $\psi_{\mathbf{k}a}\rightarrow \psi_{\mathbf{k}b}$ and $n \rightarrow -n$, the zero mode wave function on sublattice-a also hold on sublattice-b.
\section{Derivation of the topological current}
This appendix detailed the topological current in the presence of extended complex VBS order parameter,
\begin{equation}
\vec{\phi}=(Re\Delta,Im\Delta, h).
\end{equation}
We consider the Lagrangian
\begin{equation}
L=\bar{\psi}i\gamma^{\mu}(\partial_{\mu }-i\gamma _{5}a_{\mu})\psi -\bar{
\psi}\left(\sum_{k=1}^{3}\Gamma_{k}\phi_{k}\right)\psi ,
\end{equation}
where we assume $h>0$, and
\begin{eqnarray}
\Gamma _{1}=\beta \beta _{1}=\tau _{1}\sigma _{3},\\
\Gamma_{2}=\beta \beta _{2}=\tau _{2}\sigma _{3},\\
\Gamma _{3}=\beta \mathbb{I} =\tau_{0}\sigma _{3}.
\end{eqnarray}
We should note that the generalized
order parameter breaks chiral symmetry. Integrating over fermion fields, the effective Euclidean action $S_{eff}^E[\vec{n},a_{\mu }]$ is obtained via
\begin{equation}
\int d\bar{\psi}d\psi e^{-\int d^{3}xL(\bar{\psi},\psi ,\vec{n})}=e^{-S_{eff}^E[\vec{n},a_{\mu}]},
\end{equation}
the current is defined in terms of $J^{\mu }=\delta S_{eff}^E/\delta a_{\mu
}|_{a_{\mu }=0}$, Again, using the $1/m$-perturbation expansion \cite{weigman},
\begin{equation}
J^{\mu }=\frac{\delta S_{eff}^E}{\delta a_{\mu }}=-\text{Tr}\{\Gamma
_{5}\gamma ^{\mu }\mathcal{\mathcal{G}}_{0}^{-1}[\mathcal{\mathcal{G}}
_{0}^{-1}(\mathcal{\mathcal{G}}_{0}^{-1})^{\dag }]^{-1}\},
\end{equation}
where $\mathcal{\mathcal{G}}_{0}^{-1}=$ $i\gamma ^{\mu }\partial _{\mu
}-\Gamma _{k}\phi _{k}$. The large-$\phi $ expansion lead to the current
\begin{equation}
J^{\mu }=-\sum_{l=0}^{\infty }\text{Tr}[\Gamma _{5}\gamma ^{\mu }\frac{
i\gamma ^{\nu }\partial _{\nu }-\Gamma _{k}\phi _{k}}{-\partial ^{2}+\phi
^{2}}\left( \frac{-i\Gamma _{k}\gamma ^{\lambda }\partial _{\lambda }\phi
_{k}}{-\partial ^{2}+\phi ^{2}}\right) ^{l}].
\end{equation}
It's easily check that both $l=0$ and $l=1$ term vanish and the dominant term is
\begin{eqnarray}
J^{\mu } &=&\text{Tr}[\Gamma _{5}\gamma ^{\mu }\frac{\Gamma _{k}\phi _{k}}{
-\partial ^{2}+\phi ^{2}}\left( \frac{-i\gamma ^{\lambda }\Gamma
_{k}\partial _{\lambda }\phi _{k}}{-\partial ^{2}+\phi ^{2}}\right) ^{2}]\notag \\
&=&\int\frac{d^3k}{(k^2+\phi^2)^3}\text{Tr}(\gamma^{\mu}\gamma^{\nu}\gamma^{\lambda}\Gamma_5\Gamma_a\Gamma_b\Gamma_c)
\phi^{a}\partial _{\nu }\phi^{b}\partial _{\lambda }\phi^{c}\notag \\
&=&
-\frac{1}{8\pi}\epsilon^{\mu \nu \lambda }\epsilon ^{abc}\hat{\phi}
^{a}\partial _{\nu }\hat{\phi}^{b}\partial _{\lambda }\hat{\phi}^{c},
\end{eqnarray}
where $\hat{\phi}_i=\phi_i/\phi$, and the topological number is defined as $Q=\int d^{2}xJ^{0}$.
\end{appendix}

\end{document}